\documentclass{PoS}
\usepackage{amsmath, amsthm}

\usepackage{float}
\restylefloat{figure}

\newtheorem{theor}{Theorem}
\newtheorem{defin}{Definition}

\title{
Multistep Methods for Lattice QCD Simulations}

\ShortTitle{
Multistep Methods for Lattice QCD Simulations}

\author{\speaker{Dmitry Shcherbakov} and Matthias Ehrhardt\\
        Bergische Universit\"{a}t Wuppertal, 
Angewandte Mathematik und Numerische Analysis, Germany\thanks{The speaker was supported by the Marie Curie Initial Training Network STRONGnet 
\textquotedblleft Strong Interaction Supercomputing Training Network \textquotedblright }\\
        E-mail: \email{shcherbakov@math.uni.wuppertal.de},
		        \email{ehrhardt@math.uni.wuppertal.de}}
\abstract{

It is well-known that molecular dynamics integrators, which are used 
for lattice quantum chromodynamics (QCD), suffer from instabilities 
and possess a rather low order of the accuracy. 
Hence, it is highly desirable to construct a
new class of geometric integrators, that overcomes these instability problems
 and increases the order of  accuracy  without increasing  remarkably the computational costs.
 
In this paper we consider for this purpose multistep methods and 
give an overview of known results to systematize important knowledge for such methods 
being the right choice for lattice QCD simulations. 
At the end we try
to answer the question: can multistep method be used as molecular dynamic integrators 
and what might be the advantage of it.

}

\FullConference{ The XXIX International Symposium on Lattice Field Theory - Lattice 2011\\
July 10-16, 2011\\
Squaw Valley, Lake Tahoe, California}

\begin{document}

\section{Introduction}
In this paper we will give a short introduction to symplectic multistep integrators,
review some recent results and comment on the possible later application in lattice QCD computations.

 A \textit{geometric integrator} is a numerical method that preserves the geometric properties of
 the exact flow of an autonomous ordinary differential equation (ODE)
\begin{equation} \label{eq:ode}
y'=f(y), \qquad y\in\mathbb{R}^{p}.
\end{equation}
Especially, when solving numerically a Hamiltonian problem, it is of paramount importance
that the chosen scheme retains some properties of the underlying continuous problem,
like the time reversal symmetry or the area preservation property. 


We consider a compatible \textit{linear multistep methods} (LMM) with $k$ steps
of the form
          \begin{equation} \label{eq:mult}
           \sum_{j=0}^k\alpha_{j}\,y_{n+j}
           = h \sum_{j=0}^k\beta_{j}\,f(y_{n+j}),
\end{equation}
where $h=\Delta t$ denotes the time step of the grid $t_{j}=t_ {0}+jh$, 
and $\alpha_{j}$, $\beta_{j}$ are real parameters,
 $\alpha_{j}\neq0$, and $|\alpha_{0}| + |\beta_{0}|>0.$ 
For an application of \eqref{eq:mult} we need an initial value $y_0=y(t_{0})$ as well as
starting approximations $y_{1},\dots,y_{k-1}$ to
$ y(t_{0}),\ldots,y(t_{k-1})$, that are usually obtained by Runge-Kutta methods.

We recall that the LMM \eqref{eq:mult} is of \textit{order $s$} if and only if (cf.\ \cite{hairer}) 
\begin{equation} \label{eq:order}
           \sum_{j=0}^k\alpha_{j}=0,\quad
           \sum_{j=0}^k\alpha_{j}j^\ell
           = \ell \sum_{j=0}^k\beta_{j}j^{\ell-1},\quad 1\le \ell\le s,
           \qquad     
           \sum_{j=0}^k\alpha_{j}j^{s+1}
           \neq (s+1)\sum_{j=0}^k\beta_{j}j^{s}.
\end{equation}
Furthermore, the multistep method \eqref{eq:order} has the two \emph{characteristic polynomials} 
\begin{equation} \label{eq:poly}
\rho\left(\xi\right)=\sum^{k}_{j=0} \alpha_{j}\,\xi^{j},\qquad
\sigma\left(\xi\right)=\sum^{k}_{j=0} \beta_{j}\,\xi^{j}, 
\end{equation}
and the LMM \eqref{eq:mult} is called \textit{irreducible} 
if these polynomials \eqref{eq:poly} have no common roots.

Since the stability analysis for LMMs was difficult to establish
Dahlquist \cite{dahl2} suggested to consider instead so-called one-leg methods
that only need one evaluation of the forcing function $f$. 
%
The \emph{one-leg method} (OLM) associated to the  
multistep method \eqref{eq:mult} is defined by
\begin{equation}\label{eq:onel}
         \sum^{k}_{j=0}\alpha_{j}\,y_{n+j}= h f\left(\sum^{k}_{j=0}\beta_{j}\,y_{n+j}\right),
         \end{equation}
where we assume the \textit{normalization condition} 
$\sigma(1)=\sum_{j=0}^k\beta_{j}=1$.
%
The LMMs \eqref{eq:mult} and OLMs \eqref{eq:onel} are closely related. 
Let $\tilde{y}$ denote the sequence of approximated values obtained from a 
LMM \eqref{eq:mult} 
and $\hat{y}$ be the sequence of approximated values obtained 
from a OLM \eqref{eq:onel}, then we have
\begin{equation*}
\hat{y}_{j} = \sum^{k}_{j=0}\beta_{j}\,\tilde{y}_{n+j}.
\end{equation*}
Hence the analysis of the stability for LMMs can be reduced to  
the stability analysis of OLMs.

In the sequel we focus on the case that the ODE \eqref{eq:ode} is a \textit{linear autonomous Hamiltonian system}, i.e.\ $p=2n$ (where $n$ is the number of degrees of freedom in mechanics) and 
\begin{equation}\label{eq:hamil}
y'=J^{-1}\nabla H(y), \qquad y \in \mathbb{R}^{2n},\qquad
\text{where}\quad J= \begin{pmatrix}0 & I_{n}\\-I_{n} & 0\end{pmatrix}.
\end{equation}
Here, $H:\mathbb{R}^{2n}\to\mathbb{R}^{1}$ denotes a smooth Hamiltonian function. 
Now we can rewrite the LMM \eqref{eq:mult} 
\begin{equation}\label{eq:multham}
\sum\limits^{k}_{j=0} \alpha_{j}\,y_{n+j}
=h\sum\limits^{k}_{j=0} \beta_{j}\,J^{-1}\nabla H(y_{n+j}). 
\end{equation}

\section{Area-Preservation and Time-Reversibility of Multistep Methods}
A numerical integrator is used in a molecular dynamics step of the 
Hybrid Monte Carlo algorithm
has to satisfy an area-preservation property, 
which follows from the symplecticity of the numerical method, 
and a time-reversibility property, which is an extension of the symmetry.

The \textit{symmetry} is fulfilled if the coefficients of a 
  scheme \eqref{eq:mult} (or \eqref{eq:onel}) satisfy the relations
\begin{equation*}
\alpha_{k-j}=-\alpha_{j},\qquad\beta_{k-j}=\beta_{j}\quad\text{for all}\; j=0,1,\dots,k, 
\end{equation*} 
i.e.\
\begin{equation*}
 \rho(\xi)\equiv-\xi^{k}\rho(1/\xi),\qquad\sigma(\xi)\equiv \xi^{k}\sigma(1/\xi).
\end{equation*} 
They are also charaterized by having an odd number of time steps, i.e.\ $k=2\nu -1$, $v=1,2,\dots$.
This means that the numerical solutions satisfy the following reversibility requirement: 
whenever $y_{n},\dots ,y_{n+k}$ satisfy the relation
\eqref{eq:mult} (or \eqref{eq:onel}), $y_{n+k},\dots ,y_{n}$ satisfy \eqref{eq:mult} with $h$ replaced 
by $-h$. From this it follows that symmetric multistep methods are \textit{time-reversible}.

 \begin{defin}
A mapping $g:\mathbb{R}^{2n}\to\mathbb{R}^{2n}$ 
is called symplectic (with respect to $J$) 
if
\begin{equation} \label{eq:s}
\left[\frac{\partial g(y)}{\partial y}\right]^{\top}J\left[ \frac{\partial g(y)}{\partial y}\right]
\equiv J.
\end{equation}
 \end{defin}

It is well-known, that the solution of the ODE \eqref{eq:hamil} at any fixed time $t^*$,
regarded as a function on the initial data $y(0)\in\mathbb{R}^{2n}$,
(so-called \textit{phase flow}) is a \textit{symplectic mapping}.
Hence, it is a natural task to seek for 
numerical methods that retain this property
(in a sense to be specified later).

In the literature there exist (at least) two definitions for symplectic LMMs.
Eirola and Sanz-Serna \cite{ss} considered the transformation 
$g:\mathbb{R}^{2nk}\to\mathbb{R}^{2nk}$, $Y_{\ell}\mapsto Y_{\ell+1}$, where 
$Y_{\ell}=(y_\ell^\top,\dots,y_{\ell+k-1}^\top)^\top$,
that is associated to the LMM \eqref{eq:multham}, and obtained the following positive result.
\begin{theor}[\cite{ss}] 
Assume that the OLM is symmetric and irreducible.
Then the corresponding mapping $Y_{\ell}\mapsto Y_{\ell+1}$ is symplectic with respect to 
the matrix $\Lambda\otimes J$,
where $\Lambda=(\lambda_{ij})$ is given by
\begin{equation*}
\lambda_{ij}=\sum_{m\ge0} (\alpha_{i+m}\beta_{j+m}+\alpha_{j+m}\beta_{i+m}),\quad i\ge0,\;j\le k.
\end{equation*}
\end{theor}
But here the symplecticity is regarded 
with respect to another skew symmetric matrix $\Lambda\otimes J$.
We note that this symplecticity is equivalent to the preservation of quadratic first integrals, cf.\ \cite{bosco94}.
For example, Ge and Feng \cite{gefeng88} showed that the standard second order \textit{leap-frog scheme}
\begin{equation*}
y_{n+2}-y_n=2h \,J^{-1}\nabla H\bigl(y_{n+1}\bigr)\quad\text{is symplectic w.r.t.}\quad 
\begin{pmatrix}
          0 & J_{2n}\\
	  J_{2n} & 0
         \end{pmatrix}\quad\text{with}\quad\Lambda=\begin{pmatrix}
          0 & 2\\
	  2 & 0
         \end{pmatrix}.
\end{equation*} 

Contrary, in a different second approach, Tang \cite{tang93} 
proved a negative result  for the  
\textit{step-transition operator} (underlying one-step method) introduced by Feng \cite{feng98}
$g:\mathbb{R}^{2n}\to\mathbb{R}^{2n}$
satisfying 
\begin{equation} \label{eq:tran}
\sum\limits^{k}_{j=0} \alpha_{j} g^{j}
=h\sum\limits^{k}_{j=0} \beta_{j}f\circ g^{j}, 
\end{equation}
 where $g^{j}$ stands for $j$-time composition 
$g: g\circ g \circ \dots \circ g $.
This operator characterizes the LMM  \eqref{eq:mult} as 
$y_{1}=g(y_{0}),\dots,y_{k}=g(y_{k-1})=g^{k}(y_{0}),\dots$, e.g.\
for Hamiltonian systems the LMM \eqref{eq:multham} reads
\begin{equation} \label{eq:transham}
\sum\limits^{k}_{j=0} \alpha_{j}\, g^{j}
=h\sum \limits^{k}_{j=0} \beta_{j}\, J^{-1}\left( \nabla H \right) \circ g^{j}. 
\end{equation}

This step-transition operator $g$ allows for a definition of symplecticity for LMMs:
 \begin{defin}[\cite{tang93}]
  The LMM \eqref{eq:multham} is \textit{symplectic} 
   if the operator $g$ defined by \eqref{eq:transham} is symplectic, i.e.\ \eqref{eq:s} holds
  for all Hamiltonian functions $H$ and all sufficiently small step-sizes $h$.
 \end{defin}

\begin{theor}[(Conjecture of Feng) \cite{tang93}] 
For a Hamiltonian system \eqref{eq:hamil}, any compatible LMM (with order $s\ge1$ 
of form \eqref{eq:multham} is not symplectic.
\end{theor}

Despite this negative result of Theorem~2, 
it is known that the \textit{second order mid-point rule}
\begin{equation}\label{midpoint}
y_{n+1}-y_n=h \,J^{-1}\nabla H\bigl(\frac{1}{2}(y_{n+1}+y_n)\bigr).
\end{equation}
is a symplectic multistep method. This fact motivates to consider \textit{generalized LMMs} of the form 
 \begin{equation} \label{eq:gmultham}
 \sum\limits^{k}_{j=0}\alpha_{j}\,y_{j}
 = h \sum\limits^{k}_{j=0}\beta_{j}\,
 J^{-1}\nabla H\bigl(\sum_{l=0}^k \gamma_{jl}\,y_{l}\bigr),
 \qquad\text{with}\quad\sum_{l=0}^k \gamma_{jl}=1,\quad j=0,\dots,k.
\end{equation}
Unfortunately, also for  the scheme \eqref{eq:gmultham} the result is rather negative.
\begin{theor}[\cite{tang93}] 
For a Hamiltonian system \eqref{eq:hamil}, if a difference scheme (with order $s\ge1$ 
of the form \eqref{eq:gmultham} is symplectic, then it must be of order 2.
\end{theor}
In fact, Dai and Tang \cite{daitang88} showed that \eqref{midpoint} is the only symplectic
scheme of this form.
%
However, following the concept of \textit{$G$-stability} proposed by Dahlquist there exists
 another third way of transfering the definition of symplecticity to multistep methods, cf.\ \cite{dahl2}.

However, recall that for integrators to be suitable for molecular dynamics integration, 
it is sufficient to safisfy the following \emph{area-preservation property}.
 \begin{defin}
A mapping $g:\mathbb{R}^{2n}\to\mathbb{R}^{2n}$ 
is called area-preserving (without an orientation) 
if
\begin{equation} \label{eq:area}
\left|\det\frac{\partial g(y)}{\partial y}\right|
\equiv 1.
\end{equation}
 \end{defin}
which is a slightly weaker assumption than simplecticity.

\section{Numerical Experiments}

We consider a model of a \textit{simple harmonic oscillator (SHO)}  with the Hamiltonian 
\begin{equation} \label{eq:h1}
H(p,q)= \frac{1}{2}p^2 + \frac{1}{2}w^2q^2 
\end{equation} 
 to investigate the stability
 behavior of solutions for this system
by $G$-symplectic multistep methods.

\begin{figure}[ht]
\centering
 \includegraphics[width=7cm]{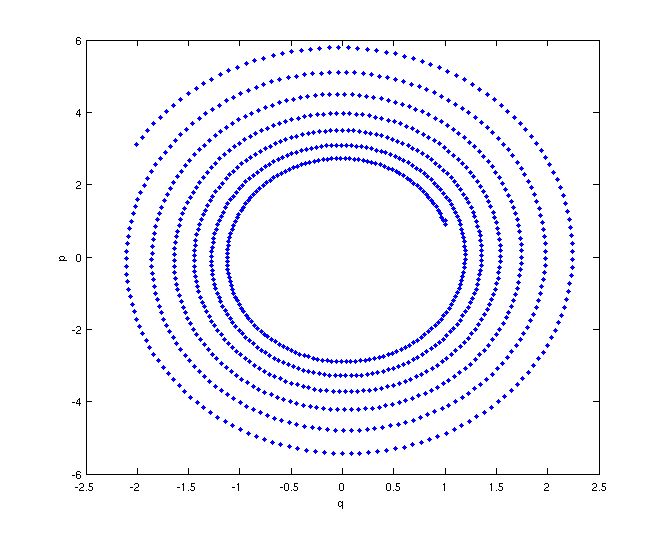} 
 \hspace{0.1cm}
 \includegraphics[width=7cm]{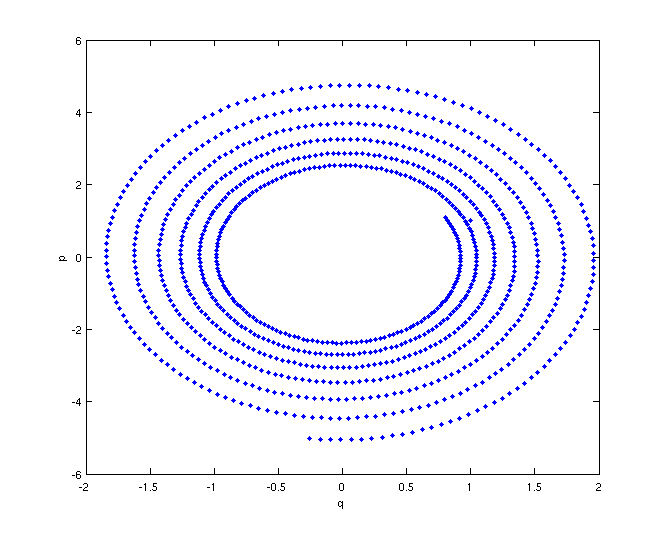} 
 \caption{Solutions of the SHO; explicit Euler method with step size $h=0.1$, initial value $(p_{0},~q_{0}) =(0,1)$;
implicit Euler method with step size $h=0.1$, initial value $(p_{0},~q_{0}) =(0,1)$.}
\label{fig:1}
\end{figure}
 
We choose the three $G$-symplectic methods and show the corresponding numerical solution of the SHO problem \eqref{eq:h1}.
The first method is a \textit{4-step explicit method}
\begin{equation} \label{eq:m1}
  y_{n+3}-y_{n+2}+y_{n+1}-y_{n} = \frac{h}{2} \left( f_{n+2}+f_{n+1} \right).
\end{equation}

The second method is a \textit{predictor-corrector method}
\begin{equation} \label{eq:m2}
\begin{aligned}
& y_{n+4}-y_{n+3}= \frac{h}{24}  \left(55 f_{n+3}-59f_{n+2}+37f_{n+1}-9f_{n} \right)~~~~&\textmd{ as a predictor} \\
& y_{n+4}-y_{n+3}= \frac{h}{24} \left(9 f_{n+4}+19f_{n+3}-5f_{n+2}+f_{n+1} \right)~~~~&\textmd{ as a corrector,} 
\end{aligned}
\end{equation}
using the explicit method to compute $f_{n+4}$ of the implicit one.
Finally, we consider a \textit{partitioned method}, 
where for each equation of the system of SHO we apply a different multistep method:
\begin{equation} \label{eq:m3}
\begin{aligned}
& y_{n+3}-y_{n+2}+y_{n+1}-y_{n}=  h \left( f_{n+2}+f_{n+1} \right),\\
& y_{n+3}-y_{n+1}= 2 h \left( f_{n+2}+f_{n+1} \right).
\end{aligned}
\end{equation}

\begin{figure}[H]
\centering
 \includegraphics[width=6.3cm]{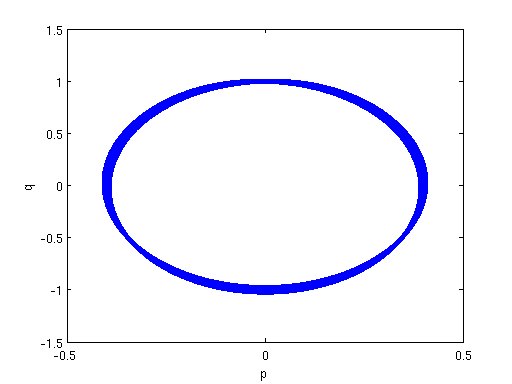} 
 \hspace{0.1cm}
 \includegraphics[width=6.3cm]{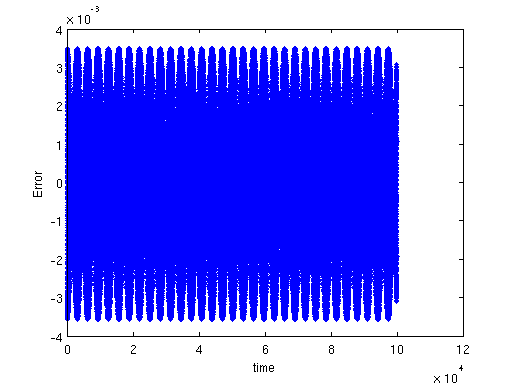}
 \caption{Solution of the SHO by the 4-step explicit method (\protect \ref{eq:m1});
with step size $h=0.1$, initial value $(p_{0},~q_{0}) =(0,1)$;
 the  error of the computed solution of the method (\protect \ref{eq:m1}) to exact solution. }
\label{fig:2}
\end{figure}

\begin{figure}[H]
\centering
 \includegraphics[width=6.3cm]{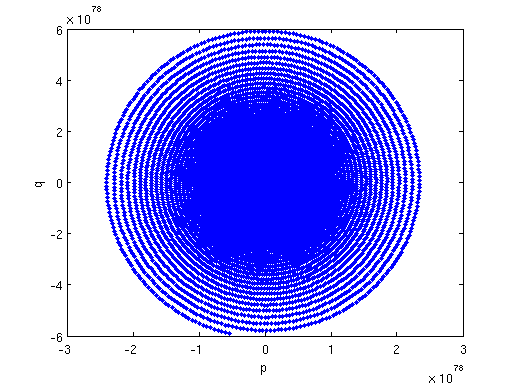} 
 \hspace{0.1cm}
 \includegraphics[width=6.3cm]{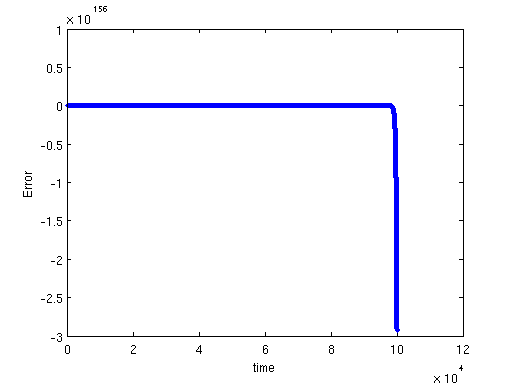}
 \caption{Solution of the SHO by the predictor-corrector method (\protect \ref{eq:m2}) with step size $h=0.1$,
 initial value $(p_{0},~q_{0}) =(0,1)$; the  error of the computed solution of the method  (\protect \ref{eq:m2}) to exact solution. }
\label{fig:3}
\end{figure}

\begin{figure}[H]
\centering
 \includegraphics[width=6.3cm]{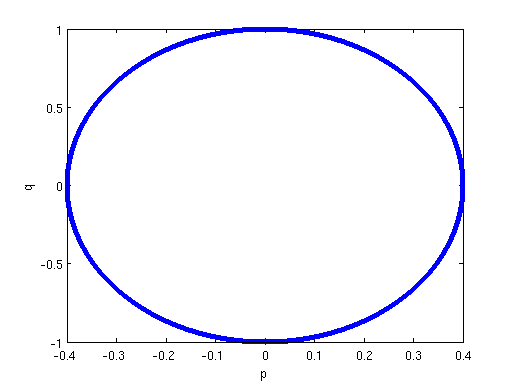} 
 \hspace{0.1cm}
 \includegraphics[width=6.3cm]{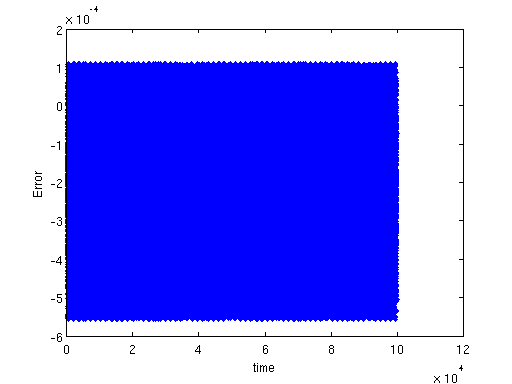}
 \caption{Solution of the SHO by the partitioned method (\protect \ref{eq:m3}) with step size $h=0.1$,
 initial value $(p_{0},~q_{0}) =(0,1)$;
 the  error of the computed solution of the method  (\protect \ref{eq:m3}) to exact solution.}
\label{fig:4}
\end{figure}

Figures~\ref{fig:2}, \ref{fig:3}, \ref{fig:4} show (from left to right) the numerical solutions obtained by the corresponding
multistep methods \eqref{eq:m1} - \eqref{eq:m3} and their long-time behavior
 ($n=1000000$ time steps).

The method \eqref{eq:m1} shows quite stable behavior for the numerical solution of the SHO, but further considerations show that increasing the 
number of steps lead us to oscillations and symplecticity of the solutions will be destroyed. But for the short term problems the method behavior
is suitable. 
The method \eqref{eq:m2} collapses after some time and the error of this method grows exponentially, but still for short time period 
it gives the proper results and the symplectic property is satisfied. 
The last method \eqref{eq:m3} yields the best results, its numerical solutions 
behave symplectic even during long time, 
it conserves the energy of system properly and this class of partitioned multistep method,
according to \cite{hairer}, gives the correct results even for the long time integrations.

\section{Conclusion and Outlook}

In spite of the collected negative results, 
the numerical experiments showed that the 
short-time behavior of the 
 multistep methods is rather promising. 
%
The main advantage of multistep method is that the high-order version of such methods can be easily obtained by one function evaluation
per  time step and it, consequently, will increase the accuracy of computations. 

Despite these predominant negative statements, we recall that in lattice QCD simulations
one solely needs an area-preserving integrator which is a slightly weaker assumption
than the discussed symplecticity.
In a forthcoming paper we will investigate, following an idea of Hairer \cite{hairer2000},
a \textit{projected LMM} that conserves the Hamiltonian and hence makes the \textit{acceptance step} in the hybrid Monte Carlo simulations obsolete.
This feature is especially interesting for small lattice spacings.

\acknowledgments
The authors acknowledge fruitful interactions with Prof.\ Yifa Tang, Beijing, China.


\end{document}